\PassOptionsToPackage{usenames,dvipsnames}{xcolor}
\documentclass[journal]{IEEEtran}

\usepackage{easyReview}
\usepackage{amsfonts}
\usepackage{listings}
\usepackage[dvipsnames]{xcolor}
\usepackage{graphicx}
\usepackage{hyperref}

\begin{document}

\title{Efficient On-line Computation of Visibility Graphs}

\author{Delia Fano Yela,
		~Florian Thalmann,
        Vincenzo Nicosia,
        Dan Stowell,
        and Mark Sandler 

\thanks{This work was funded by EPSRC grant EP/L019981/1}}

\maketitle

\begin{abstract}
A visibility algorithm maps time series into complex networks following a simple criterion. The resulting visibility graph has recently proven to be a powerful tool for time series analysis. However its straightforward computation is time-consuming and rigid, motivating the development of more efficient algorithms. Here we present a highly efficient method to compute visibility graphs with the further benefit of flexibility: on-line computation. 
We propose an encoder/decoder approach, with an on-line adjustable binary search tree codec for time series as well as its corresponding decoder for visibility graphs.
The empirical evidence suggests the proposed method for computation of visibility graphs offers an on-line computation solution at no additional computation time cost.
The source code is available online.

\end{abstract}

\begin{IEEEkeywords}
on-line, visibility graphs, binary trees, networks.
\end{IEEEkeywords}

\section{Introduction}

\IEEEPARstart{I}{n} the last decade, several methods to map time series into graphs have been proposed under the hypothesis that, appropriate graph representations can preserve the original time series information while providing alternatives to deal with non-linearity and multi-scale issues typical of complex signals~\cite{Donner2011,NunezLGL12_ReviewVg_InTech}.
This line of research represents a bridge between nonlinear signal analysis and complex network theory, and has been successfully applied to extract meaningful information from a variety of different systems in physics~\cite{Shirazi2009,Lacasa2010}, finance~\cite{Fiedor2014,LacasaNL15_multiplex_Nature,Musumeci2017}, engineering~\cite{TorreGLBLLH17_speechLaws_Nature}, and neuroscience~\cite{Bullmore2009,Bullmore2012}.

The most notable algorithms to construct a graph from an ordered sequence of data points are either based on correlation~\cite{Mantegna1999a,Bonanno2001,Yang2008}, recurrence~\cite{Donner2010a,Donges2013,Feldhoff2013}, dependence~\cite{Marinazzo2008,Liao2011}, or visibility~\cite{LacasaLBLN08_VisibilityGraph_PNAS}.
However, the visibility algorithms proposed by Lacasa et al.~\cite{LacasaLBLN08_VisibilityGraph_PNAS, LuqueLBFL09_HVg_PRE} are amongst the most popular as they provide a deterministic and non-parametric symbolisation of a time series preserving full information of its linear and non-linear correlations. Such visibility algorithms can also effectively deal with non-stationary signals and are deemed computationally efficient. 
In consequence, visibility graphs have found numerous applications in diverse fields including image processing~\cite{Iacovacci2017,Iacovacci2018}, number theory~\cite{Lacasa2018}, finance~\cite{LacasaNL15_multiplex_Nature,Flanagan2018}, and neuroscience ~\cite{Sannino2017}.

The straightforward computation of visibility graphs presents a worst case time complexity quadratic in the length of the series. Even though such complexity should not be an issue for medium-sized series ($10^4-10^5$ points), it remains inefficient for longer time series. Therefore, faster algorithms have been proposed employing a `Divide \& Conquer' (DC) approach, reducing the average-case time complexity to $O(n \log n)$~\cite{Lan15_DC_Caos}.

Both of these approaches comprising the current existing methods to compute visibility graphs, are off-line algorithms, as they require all the data points in the time series to be available before the graph is constructed. Consequently, the integration of new data points normally requires to re-compute the visibility graph from scratch, representing a major shortcoming limiting the real-world applications of visibility graphs.

In this paper we present, to the best of our knowledge, the first on-line algorithm to compute visibility graphs efficiently.
The proposed algorithm employs an `encoder/decoder' approach by means of a binary search tree representation of the time series (or any ordered sequence of data points). 
In particular, the time series is encoded into a binary search tree that can be updated 
every time a chunk of time series is available by merging its corresponding binary search trees. 
The resulting binary search tree can subsequently be decoded into a visibility graph when required. 
This introduced flexibility comes at no significant computational cost as the presented method shares the computational complexity of the current fastest visibility algorithm (DC). 

\begin{figure}[!b]
  \centering
  \includegraphics[width=3.6in]{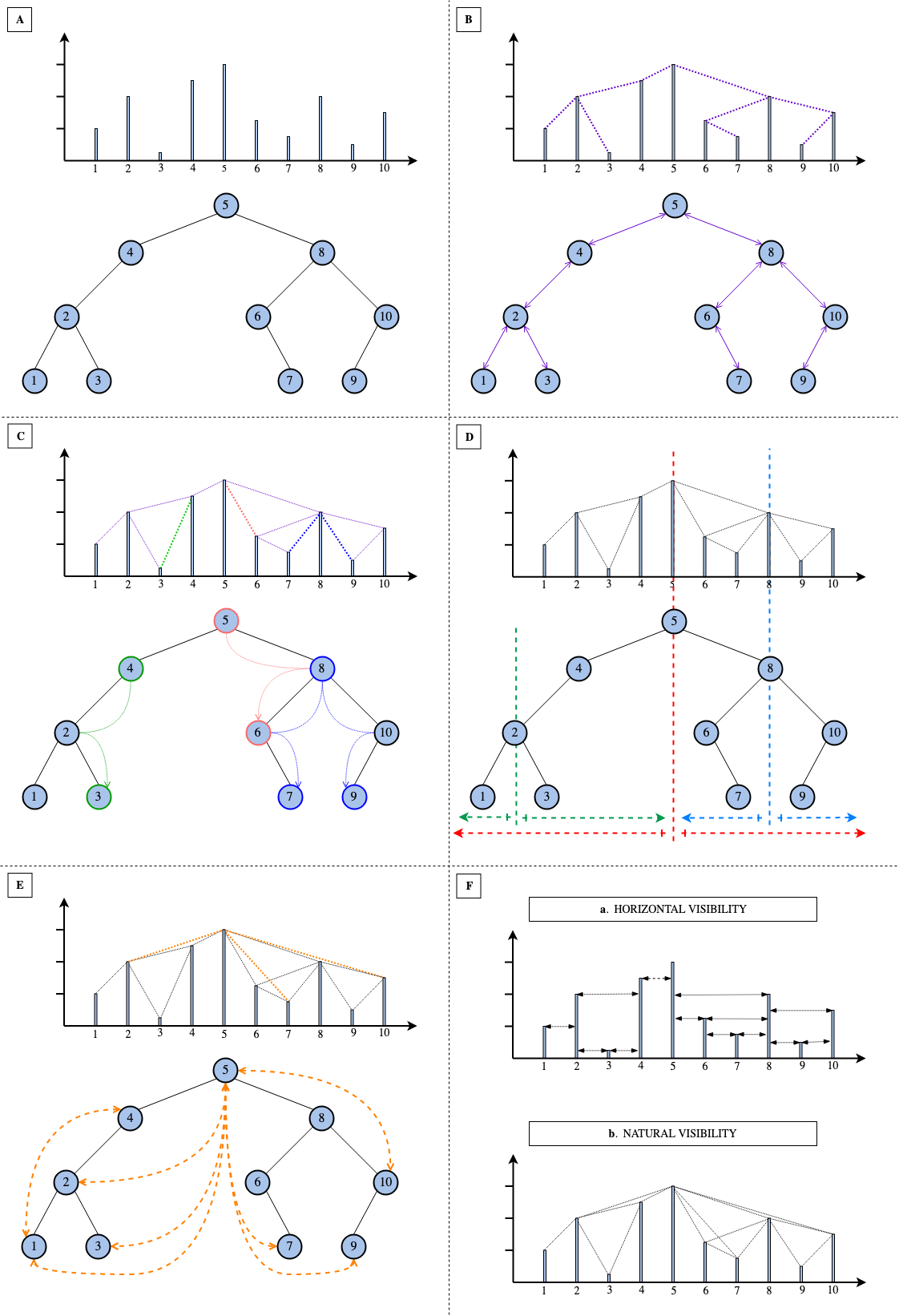}
  \caption{Representation of the different steps of the proposed algorithm for visibility graphs computation. In section A, the sample time series and its correspondent maximum binary search tree. Section B represents the connections deduced by the first connectivity rule. The second and third connectivity rules are illustrated in section C and D respectively. Section E shows the remaining checks needed to ascertain natural visibility. Finally, section F reports the horizontal and natural visibility graph associated to the original time series.}
  \label{fig:method}
\end{figure}


\section{Visibility Graphs}

A visibility graph is obtained from an ordered sequence of values by associating each datum to a node and connecting two nodes with an edge if the corresponding data points are visible from each other. 
A point $a$ is visible from the point $b$ if one can draw a straight line from $a$ to $b$ without passing underneath any intermediate points. In this paper we will consider visibility as a symmetric relation, so that the resulting visibility graphs are undirected. 

The natural visibility criterion (NV) allows the visibility line between $a$ and $b$ to take any slope, whereas the horizontal visibility criterion (HV) is restricted to horizontal lines, as shown in Figure \ref{fig:method}.f. 
More precisely, given a time series 
\[
y = f(t) 
\]
of length $n$, two points $(t_{a}, y_{a})$ and $(t_{b},y_{b})$ are said to be naturally visible if every intermediate point $(t_{c}, y_{c}) $, such that $t_{a}<t_{c}<t_{b}$, fulfills the following simple geometrical criterion:
\[
y_{c} < y_{a} + ( y_{b} - y_{a} ) \frac{t_{c} - t_{a} }{t_{b} - t_{a}}
\]
This natural visibility criterion will therefore establish the connections between nodes in the resulting natural visibility graph (NVG). 

One can analogously map a time series into a horizontal visibility graph (HVG) where two points $(t_{a}, y_{a})$ and $(t_{b},y_{b})$ are said to be horizontally visible if :
\[
y_{a}, y_{b} > y_{c} ~ \forall c ~ such ~that ~t_{a} < t_{c} < t_{b}
\]

From the definition of visibility it immediately follows that, for a set visibility criterion, the visibility graph associated to a given time series is unique. Moreover, any two subsequent data points of the time series are always connected by an edge, thus visibility graphs are connected and Hamiltonian~\cite{Lacasa2009}. In addition, visibility graphs are also invariant to re-scaling on both horizontal and vertical axes (i.e., the first point on either side of a node $i$ remains visible from $i$ no matter how far apart they are), and invariant to vertical and horizontal translations (i.e., only the relative values of point determine visibility relations).

In Figure \ref{fig:method}.f. we show both the natural and horizontal visibility criteria at work on an arbitrary time series. Notice that  horizontal visibility is a more stringent criterion than natural visibility, meaning that if two points are horizontally visible then they are also trivially visible when using the natural visibility criterion. Consequently, the horizontal visibility graph of a time series is always a sub-graph of the natural visibility graph associated to the same time series.

\section{State of the art}

A straightforward approach to compute visibility graphs consists in checking whether any of the points of the time series is visible or not from every other point. This corresponds to evaluating the visibility criteria for every pair of points in the time series. Since we consider visibility as a symmetric relation, the total number of checks needed to obtain a visibility graph of a time series of $n$ data points is equal to $n(n-1)/2$, corresponding to a $O(n^2)$ time complexity.

In the case of horizontal visibility, one can take a step further and safely assume that no point after a value larger than the current value $t_{a}$ will be horizontally visible from $t_{a}$. This observation effectively reduces the time complexity of the construction to $O(n \log(n))$ and, in the case of noisy (stochastic or chaotic) signals, it can be proved that this algorithm has an average-case time complexity $O(n)$~\cite{Lacasa2009}. Nevertheless, all pairs of points need to be checked in the case of natural visibility. From now on, this simple approach will be referred to as the basic method for both natural and horizontal visibility computation~\footnote{The original Fortran 90 implementations of basic algorithms to construct visibility graphs can be found at \url{http://www.maths.qmul.ac.uk/~lacasa/Software.html}}. 

As an improved alternative for visibility computation, Lan et al. presented a `Divide \& Conquer'  approach~\cite{Lan15_DC_Caos}. This algorithm reduces the average case time complexity of the construction of the natural visibility graph to $O(n \log(n))$ and it significantly reduces computation time for most balanced time series.

The basic idea behind the `Divide \& Conquer' algorithm is related to the horizontal visibility optimisation mentioned above. Once the maximum value $M$ of the time series is known, one can safely assume that the points on the right of $M$ will not be naturally visible from the points on the left of $M$ (the point $M$ is effectively acting as a wall between the two sides of the time series). The same argument is then applied recursively on the two halves of the time series separated by $M$, where the local maxima subsequently found at each level are connected with an edge to the maxima at the level immediately above them. From now on, this improved method will be referred to as `Divide \& Conquer' (or DC for short). 

Both the basic method and DC are off-line approaches, meaning that they require all the points of the time series to be accessible at the beginning of the computation. This rigid requirement limits the applicability of visibility graphs, specially in fields like telecommunications or finance, where there is a constant incoming flow of new data to be processed and assimilated. Moreover, in such big data scenarios, one tends to favour an initial overall high level analysis that will reveal the need for further processing. This work-flow would benefit from dynamic algorithms unlike the ones presented above.

\section{Proposed Method: Binary Search Tree Codec }

\begin{figure}
  \hspace*{-0.2in}
  \includegraphics[width=3.6in]{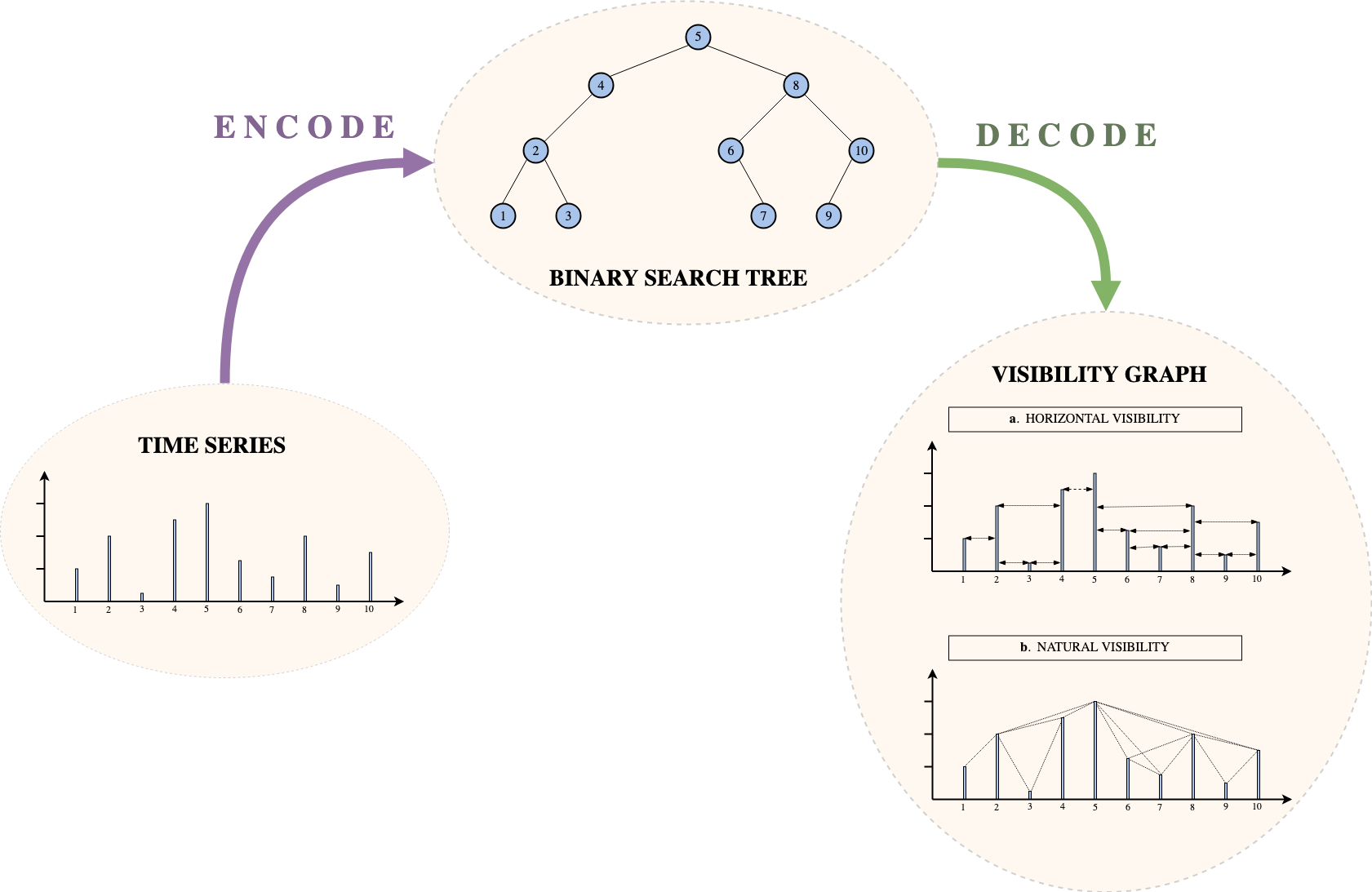}
  \caption{Illustration of the encode/decode approach of the proposed method to calculate visibility graphs. }
  \label{fig:encode}
\end{figure}

Here we propose a new method to compute visibility graphs on-line based on an encoding/decoding approach. In our method, the necessary visibility information is first encoded into an appropriately constructed binary search tree, and then successively decoded into a visibility graph when needed, as shown in Figure~\ref{fig:encode}.

\subsection{Encoding - Maximum Binary Search Tree}
The construction of a maximum binary search tree is fairly straightforward and its corresponding pseudo-code is shown in Algorithm~\ref{Build}. The first step is to sort the given time series in descending order of values, while storing the original position of each value in the time series. From now on, we will refer to the original positions as indices (i.e. $t$) and to the  values of the times series simply as values (i.e. $y(t)$). In the case of repeated values in the sequence, the first encountered index will come first while sorting. 

\renewcommand{\ttdefault}{pcr}

\renewcommand\lstlistingname{Algorithm}
\renewcommand\lstlistlistingname{Algorithms}
\def\lstlistingautorefname{Alg.}

\lstdefinelanguage{pseudo}
{
    keywords=[1]{def, class, if, else, for, not, return, in, is},
    keywordstyle=[1]{\bfseries},  
    keywords=[2]{Node, float},
    keywordstyle=[2]{\bfseries \textcolor{Peach}},     
    keywords=[3]{add, __init__, merge, buildTree},
    keywordstyle=[3]{\bfseries \color{NavyBlue}},
    keywords=[4]{value, index},
    keywordstyle=[4]{\bfseries \color{PineGreen}},
    keywords=[5]{left, right},
    keywordstyle=[5]{\bfseries \color{RedViolet}},
    keywords=[6]{append, remove, min_index, maxima_value, sign, sort_descending, getIndex},
    keywordstyle=[6]{\bfseries \color{Blue}},
    keywords=[7]{empty, null},
    keywordstyle=[7]{\bfseries \color{Gray}}                
}

Once we have a list of values sorted in descending order, together with the corresponding indices, we follow the standard procedure to build a binary search tree based on the indices. Every entry in the index list will be a node and each node has a left and right child, as shown in the data structure proposed in Algorithm~\ref{Build} (i.e., \emph{Node}). The first node of the binary tree (the one with no $parent$) is called $root$. In our case, the root will be the index of the datum corresponding to the maximum value in the time series, which is also the first entry in the index list. 

The next index, corresponding to the point with the largest value smaller than the maximum, will then be added to the tree. If its index is smaller than the root, it will become the left child of root, while if its index is larger than the root it will become the right child of root (see function \emph{add} in Algorithm~\ref{Build}). The next index to add will start off being compared to the root; if it’s smaller, it will travel to the left of the tree and, if it’s bigger, to the right. It will continue descending the tree in this manner until it finds an empty spot. We continue adding the indices in the list accordingly (see function \emph{build\_tree} in Algorithm~\ref{Build}) until there are no more data points (i.e. indices) to add. 

In the case of the sample time series in Figure~\ref{fig:method}.a, the maximum is in position 5 and will therefore be the $root$ of the binary tree. The point whose value is immediately smaller than the maximum is in position 4 (less than 5), so it will become the left child of the root. The third point in the list is in position 2, and will travel down the tree on the left-most branch (as it is smaller than both 5 and 4). The right branch of the tree is populated by the fourth point (in position 8), whose index is bigger than the $root$. In Figure~\ref{fig:method}.A one may appreciate the correlation between the time series and its associated binary tree structure. The visibility information captured by such tree may also now be more apparent. 

The time complexity of the procedure needed to encode the time series into the maximum binary search tree is $O(S + T)$ where $O(S)$ is the time complexity of sorting the series and $O(T)$ is the time complexity of the algorithm to construct the binary search tree. Sorting by comparisons is known to be $O(n \log n)$ (e.g., by using either MergeSort of QuickSort), while constructing a binary search tree costs on average $O(n \log n)$. Hence the overall average-case time complexity of the encoding step is $O(n \log n)$.

\vspace{0.5cm}
\begin{lstlisting}[caption= Pseudocode of the algorithm used to build a maximum binary search tree, 
			  captionpos=b,
		      language= pseudo, 
		      basicstyle=\fontsize{8}{10}\ttfamily, 
		      frame=single, 
		      tabsize=2,		     
		      mathescape=true,
              label = Build
		      ]
Node {
	index : float 	# x, input, argument
	value : float 	# f(x), output
	left  : Node  	# left child subtree
	right : Node	  # right child subtree
     }

def buildTree(values : {float}, indexes: {float}):

    root $ \leftarrow $ Node()

    sorted_values = sort_descending(values)
    sorted_indexes = indexes[getIndex(sorted_values)]

    for (i, v) in (sorted_indexes, sorted_values):
        root.add(Node(index = i, value = v))

    return root


def add(self : {Node}, node : {Node}):

    if  self is empty :
            self.index  = node.index
            self.value  = node.value
    else:
        if node.index < self.index:
            self.left.add(node)
        else:
            self.right.add(node)
\end{lstlisting}

\subsection{Decoding - Connectivity Rules}

The structure of the maximum binary search tree encodes sufficient information about the time series to allow to efficiently construct the corresponding horizontal visibility graph. The decoding procedure is based on the following connectivity rules, also illustrated in Figure~\ref{fig:method} :
\begin{enumerate}
	\item All the nodes connected by an edge in the maximum binary search tree are visible to each other and therefore connected in the visibility graph (Figure~\ref{fig:method}.B);
	\item Each node of the maximum binary search tree sees all the nodes in the left-most branch of the sub-tree rooted at its right child, as well as all the nodes in the right-most branch of the sub-tree rooted at its left child (Figure~\ref{fig:method}.C);
	\item The nodes of the left sub-tree of a node $i$ are not visible from the nodes of the right sub-tree of node $i$ (Figure~\ref{fig:method}.D)
\end{enumerate}

Note that, if there are no adjacent repeating amplitudes, the horizontal visibility graph is fully determined by these connectivity rules. In particular, when checking the connectivity rules, we simply skip a node if it has the same value as the current node. One can think of adjacent points with equal value as an interconnected `super node', which takes the smallest index value when `looked' from the left and the biggest index value when `looked' from the right or from above. 

Since the horizontal visibility decoding will always be fully determined by the three connectivity rules above, its time complexity is the sum of the time complexity of the rules. 
Essentially, each rule can be reduced to a series of look-ups in a binary search tree, and each look-up operation has time complexity $O(\log(n))$ in a balanced tree. These connectivity rules are applied to every node in the tree, and so the overall time complexity of decoding a horizontal visibility graph is $O(n \log(n))$. This represents a major improvement over the state-of-the-art algorithms, which  can ramp up to $O(n^2)$ in the worst case scenario.

The construction of the natural visibility graph, instead, requires the creation of some connections that are not captured by the three connectivity rules above. Hence, in this case we need to perform additional visibility checks (Figure \ref{fig:method}.E). In particular, for each node $i$ we must check the natural visibility criterion with each node in the sub-tree rooted at the right child of $i$ and with each node in the sub-tree rooted at the left child of $i$. These additional checks do not modify the average-case time complexity (which remains $O(n \log n)$, but the worst-case scenario still depends on the actual structure of the time series, and yields a time worst-case time complexity $O(n^2)$ for monotonically increasing or decreasing time series.

\begin{figure}[t]
\hspace*{-0.1in}
  \centering
  \includegraphics[width=3.6in]{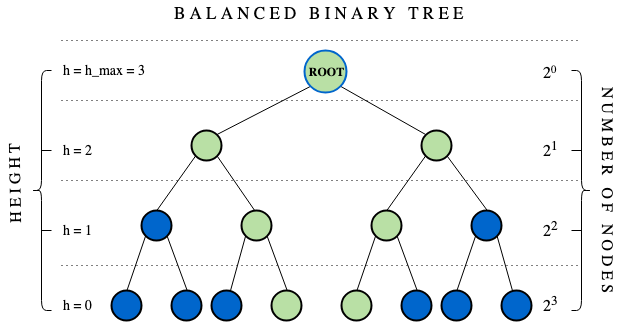}
  \caption{Representation of a perfectly balanced tree of height 4. The nodes in green are visible to the root and this visibility can be deduced by the proposed decoder (i.e. the connectivity rules). The number of nodes at each height in a balanced tree can always be expressed in base 2. }
  \label{fig:balanced}
\end{figure}

\subsection{Time Complexity}

In order to determine the time complexity of the proposed method, we will follow the standard procedure by considering the worst-case and average-case scenarios. In both scenarios, the time complexity of the encoding stage is determined by the time complexity of the sorting algorithm used, which in general is $O(n \log(n))$, and of the construction of the binary search tree, which is $O(n \log n)$. So in both cases encoding into a binary search tree costs $O(n \log n)$. 

Decoding into a horizontal visibility graph is made through the three rules explained in Figure~\ref{fig:method}B-D, which require only a visit of the binary search tree (with time complexity $O(n)$). Hence, the overall time complexity of encoding and decoding into a horizontal visibility graph  is $O(n \log(n))$. 

The worst case for decoding into a natural visibility graph is that of monotonically increasing, monotonically decreasing, or constant series, whose corresponding binary search trees degenerate into a line. In this case, the second and third connectivity rules are trivial, leaving only the first rule and the additional natural visibility checks. More precisely, if the tree is a line we need to check the natural visibility among $(n-1)(n-2)/2$ pairs of nodes, while the visibility of the remaining $(n-1)$ pairs of nodes is determined by the first connectivity rule. Even though this requires $(n-1)$ checks less than the basic implementation (which requires $n(n-1)/2$), the time complexity will still be $O(n^2)$ for the worst case scenario.

For the average case we assume the maximum binary search tree to be balanced. This means that the connectivity rules of the decoder will significantly reduce the overall number of visibility checks. If we consider a perfectly balanced binary tree as shown in Figure~\ref{fig:balanced}, the inner left branch of the right sub-tree and the inner right branch of the left sub-tree of a node are visible to the parent node. These are represented in green in Figure~\ref{fig:balanced} where the root is the parent node. This means that the visibility between the root and all the rest of nodes (the ones in blue) is unknown and needs to be checked.  

Therefore we can deduce that the number of remaining visibility checks for the root in a balanced tree of height $h_{\rm max}$ is equal to $2^{h_{\rm root}+1} - 1 - 2h_{\rm root}$ , where $2^{h_{\rm root}+1} - 1$ is the total number of nodes below the root while $2h_{\rm root}$ is the number of nodes whose visibility can be deduced by the three decoding rules (green nodes). Notice that the height of the root $h_{\rm root}$ corresponds to the maximum height of the balanced tree $h_{\rm max}$.
The same reasoning applies to all the other nodes. More precisely, for a node at height $h$, there will be $(2^{h+1} - 1 - 2h)$ remaining visibility checks to be performed. 

In order to calculate the total number of remaining visibility checks, one needs to multiply the individual expression above by the number of nodes at that height $2^{h_{max}-h}$ and sum across all heights where the checks are needed (all except the last two).  Therefore, one can express the total number of remaining natural visibility checks in a perfectly balanced binary tree as follows:
\[
\sum_{h=2}^{h_{\rm max}} 2^{h_{\rm max}-h}\big[ 2^{h+1} - ( 2h +1 ) \big] = 
\]
\[
2^{h_{\rm max}} \big[ 2 ( h_{\rm max} - 1 ) - \sum_{h=2}^{h_{\rm max}} h2^{1-h} - \sum_{h=2}^{h_{\rm max}} 2^{-h} \big]
\]
Since the maximum height of a balanced tree with $n$ nodes is $h_{\rm max} =\log_2(n)$, the total number of operation is dominated by the first term of the expression above,
\[
2^{h_{\rm max}} 2 ( h_{\rm max} - 1 ) = 2n ( \log_2(n) -1 )
\]
while the remaining terms will only introduce logarithmic corrections. In conclusion, the time complexity of the decoding for natural visibility graphs is on average $O(n\log(n))$. 

The proposed method has the same average-case time complexity than the DC algorithm, thus improving on the original basic algorithm for both horizontal and natural visibility graphs. In the Experiment section below we will see that in practice our algorithm out-competes the basic algorithm and performs as well as the DC approach, with the additional property of allowing for on-line assimilation of new data points.

\vspace{0.5cm}
\begin{lstlisting}[caption= Pseudocode of the proposed algorithm to merge two binary trees defined by their root (class Node). The input is a list of roots to be merged., 
			  captionpos=b,
		      language= pseudo, 
		      basicstyle=\fontsize{8}{10}\ttfamily, 
		      frame=single, 
		      tabsize=2,		     
		      mathescape=true,
              label = Merge
		      ]
def merge(input:{Node}):

    if input is empty: return null

    r    $ \leftarrow $ min_index(maxima_value(input))
    pool $ \leftarrow $ input $\setminus$ {r}

    pool.append(r.left, r.right)

    for n in input $\setminus$ {r} :
        
        for c in [n.left, n.right] :
        
            if sign(n.index - r.index) 
                $\neq$ sign(c.index - r.index):
                
                pool.append(c)
                n.remove(c)

    return Node(
        index = r.index,
        value = r.value,
        left  = 
          merge({p | p $\in$ pool, p.index < r.index }),
        right = 
          merge({p | p $\in$ pool, p.index > r.index }))
\end{lstlisting}

\section{On-line visibility graphs: merging binary trees}

Every time a node is added to an existing binary search tree, it essentially `travels' down the tree, going left if smaller and right if larger, until it finds an empty space (see pseudocode function $add$ in Algorithm~\ref{Build}). Therefore when a node is added to an existing binary tree there is no need to recalculate the tree structure from scratch. 
Due to the fact that the proposed encoder is a binary search tree, there is a possibility to efficiently update it on-line.

\begin{figure*}[t]
  \centering
  \includegraphics[width=1.6\columnwidth]{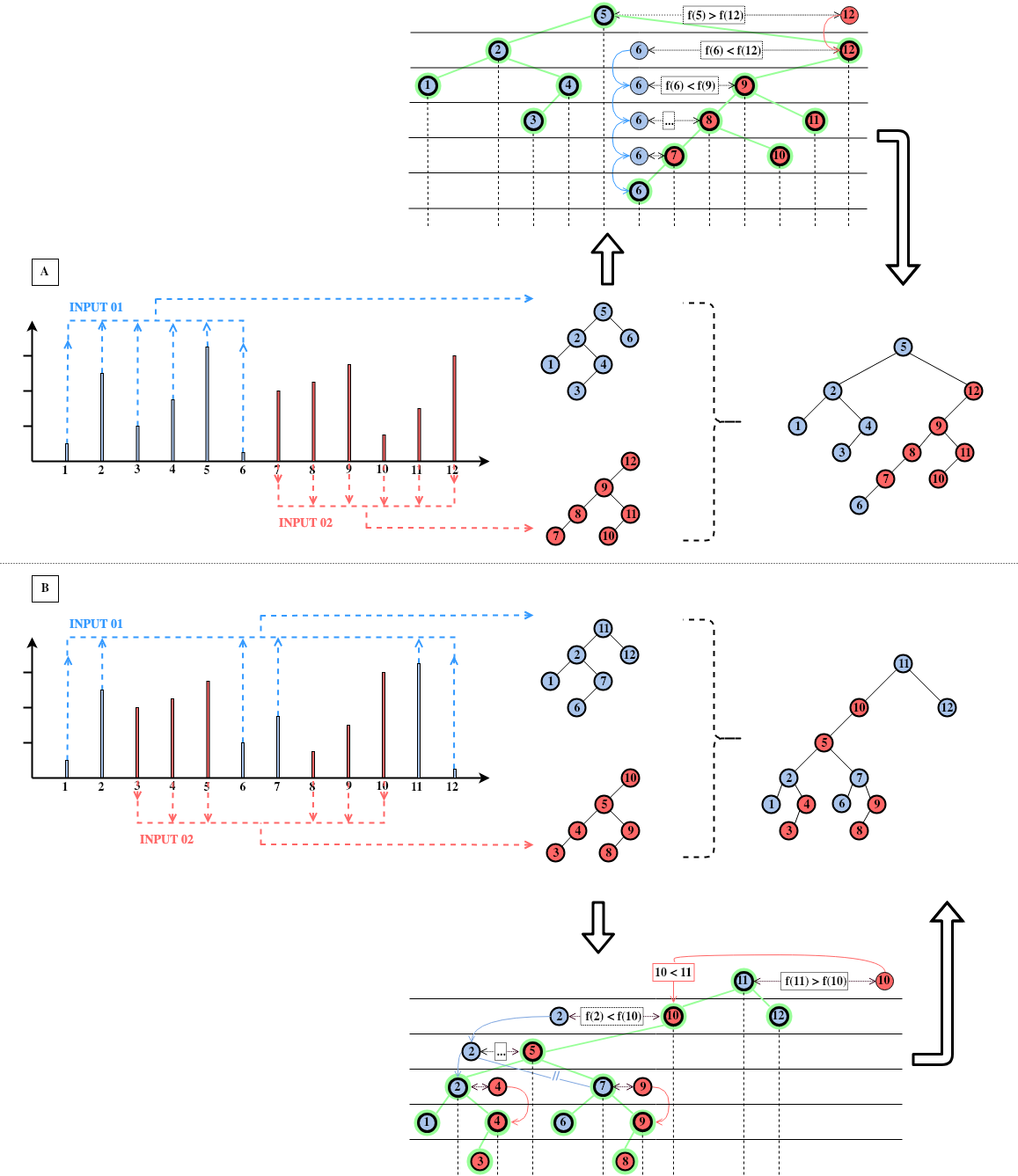}
  \caption{Visual representation of the proposed method to merge two maxima binary trees, covering both append (A) and insert (B) operations. This corresponds to an on-line scenario where a new batch (red) needs to be incorporated to an existing structure (blue). }
  \label{fig:merge}
\end{figure*}

Given a time series and its correspondent binary search tree, we would like to integrate new data points in the tree structure without recomputing it from scratch. 
One could process the points of the newly available batch of data individually and include them in the existing tree structure by comparing both values and indices.
However, other than being a time consuming approach for large numbers of points, processing points individually fails to include useful information of both the batch and the current tree structure. 
As an example, in Figure \ref{fig:merge}.A, all the nodes in the batch to be added (red nodes) have larger indices than the nodes in the current tree structure (blue nodes), and so larger indices than the current root. This means, all the nodes in the batch will populate the right side of the resulting tree. If the nodes are treated individually, this information will be overlooked producing an inefficient algorithm.

Therefore, we propose to take a different approach by treating the new batch of points as an entity. More precisely, we propose to compute the binary search tree of the new nodes and $merge$ it with the previous tree structure as illustrated in Figure \ref{fig:merge}. 
In this way, if all the new nodes indices are larger than the current root, one can include such information and produce an optimised algorithm, where potentially only one comparison is needed to merge the current with the batch tree.
This is the case for real-time incoming data, as the batch's nodes always have larger time values (indices) than the previous points in the time series.

Furthermore, the proposed merge approach covers both append and insert operations, illustrated in Figure \ref{fig:merge}.A and Figure \ref{fig:merge}.B respectively. In terms of time series representation, this means one could update the binary tree codec with observations that happened later in time or with a higher time resolution. This novel introduced flexibility for visibility computation, opens the door to new applications such as big data or audio applications where the sampling rate may vary at different analysis stages.

In order to merge two trees, we propose to compare them by levels, increasing depth at every recursion of the $merge$ function outlined in Algorithm~\ref{Merge}. 
The comparison happens in two steps: firstly the node values at a level are compared to determine which node will occupy that location in the resulting tree; secondly, the node indices are compared to determine which direction the rest of the nodes will travel down in depth.

Following the construction of the proposed binary search tree, the node with larger value will be chosen and the rest of the nodes will travel left if their indices are smaller than the chosen one and right otherwise. 
The nodes to be compared are the children of the chosen node with the nodes from the previous level that were not chosen; starting of by comparing the two roots of the trees to be merged. 
The $merge$ algorithm is illustrated step by step at the top and bottom of Figure~\ref{fig:merge}.

In the example in Figure \ref{fig:merge}.A, the blue and red tree are to be merged. Initially, the blue root is compared to the red root. Since the blue root has a larger value than the red root, it will be chosen to take that position in the resulting tree (i.e. the root of the resulting tree). The red root will then travel down the right branch as it index is larger than the chosen blue root, leaving the left branch of the chosen blue root untouched. 

Consequently the right blue child is to be compared with the red root. In this case, the red root has a larger value and so it will take the right branch position in the resulting tree. Now is the turn to the right blue child to descend down the red tree. Since the blue child happens to be the lowest value in the series, it will just descend layers following the binary search tree rules until is reaches an empty spot.

Usually, as one may observe in Figure \ref{fig:merge}, the children of the nodes that travel down in depth are not included in the level comparison. 
However, when new data is to be inserted to the existing series, the child of the node traveling down could have an index corresponding to the other branch of the resulting tree. In this case, the connection between the node and that child will be broken thereafter.
For example, in Figure \ref{fig:merge}.B., this situation takes place in layer 3, where Node 7, the child of Node 2 belongs on the right branch of Node 5 unlike its parent.

\begin{figure*}[h]
  \includegraphics[width=\textwidth]{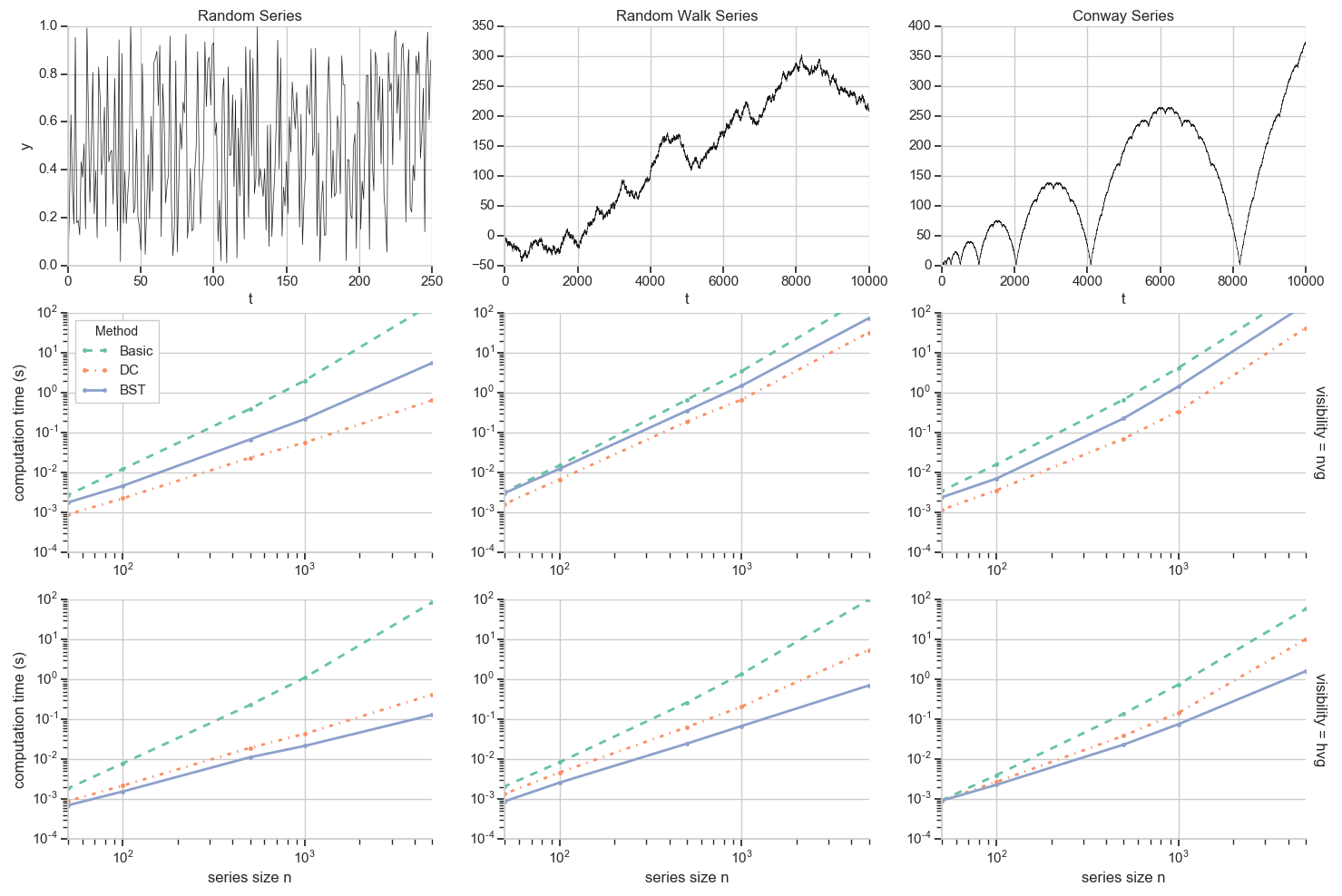}
  \caption{Computation time of the natural visibility graph (nvg, second row) and horizontal visibility graph (hvg, third row) of different time series (examples on first row) using the current visibility algorithms: Basic, Divide \& Conquer (DC), and the proposed binary search tree (BST) method. Each point at every series size is the mean of the computation time for 10 series of that size. }
  \label{fig:series}
\end{figure*}

\begin{figure*}[h]
  \includegraphics[width=\textwidth]{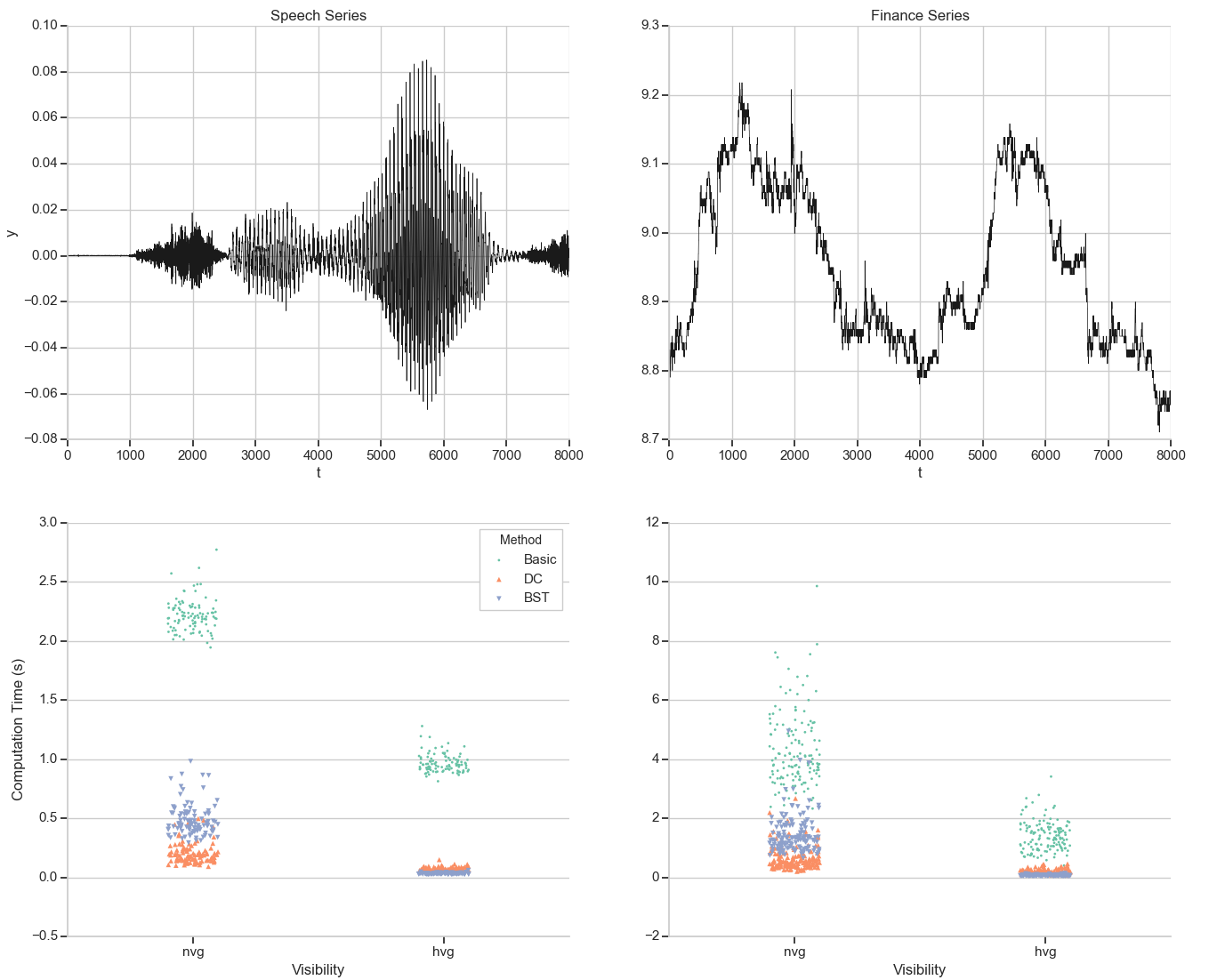}
  \caption{Current and proposed visibility algorithms computation time for 100 speech and finance time series of 1000 points. The speech time series are sampled from the training TIMIT dataset \cite{Garofolo93_TIMIT_NASA}. The finance time series corresponds to the 2013 quarterly data used in \cite{musmeci2017multiplex}. }
  \label{fig:speech_finance}
\end{figure*}

\begin{figure}[h]
  \includegraphics[width=\columnwidth]{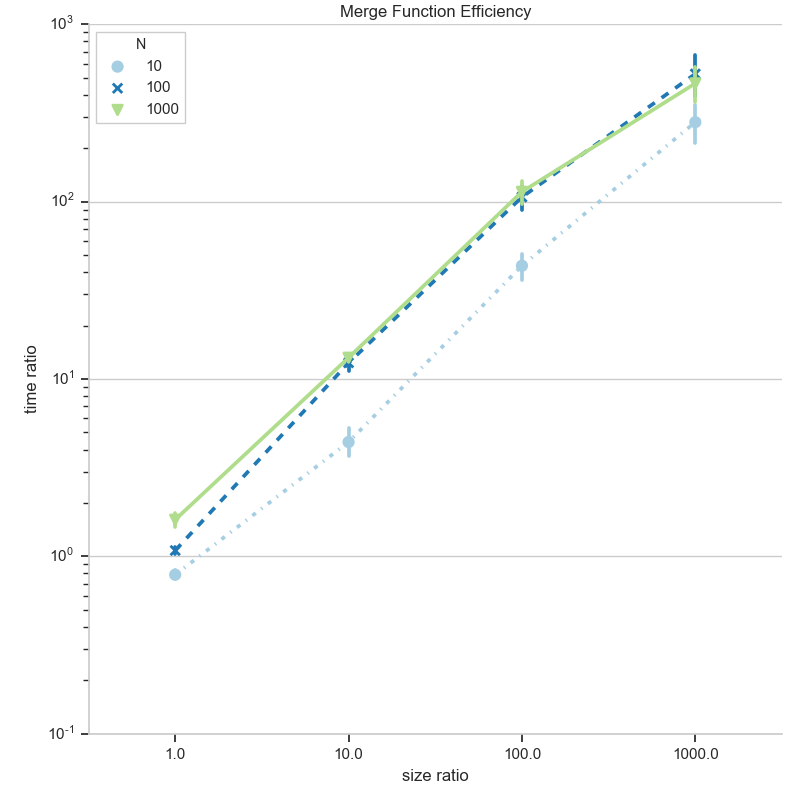}
  \caption{Given a random time series (size $L$) and a batch of new random points (size $N$) to be added to it, this plot shows the advantage, in terms of computation time, of the proposed on-line approach versus the off-line alternative. The proposed method computation time is the time it takes to build the maxima tree of the new points and merge it with the existing time series tree (i.e. $t_{on-line}$). The off-line alternative computation time is the time it takes to build a new maxima binary tree from scratch including the new points to the time series (i.e. $t_{off-line}$).  The time ratio is the $log$ scale of $t_{off-line} / t_{on-line}$, how much quicker the proposed method is. The size ratio is $L/N$, how much bigger the time series is compared to the batch to be added.  Both append and insert scenarios are represented here, 10 random cases of each were computed. The point in the graph is the mean of these 20 cases and its uncertainty is captured by the error bars.}
  \label{fig:merging}
\end{figure}

\section{Numerical Experiments}

In this section we present empirical results in order to show how the proposed visibility algorithm compares to the state of the art. All the code related to this paper and necessary to run the following experiments is implemented in Python 2.7 and freely available online \footnote{Available at \url{https://github.com/delialia/bst}}. The machine used in the simulations is an early 2015 MacBook Pro Retina with a 2.9GHz Intel Core i5 processor and 16GB of RAM. 

To put the presented algorithm into context~\cite{Lan15_DC_Caos}, in Figure~\ref{fig:series} we report the computation time needed by current visibility algorithms on different synthetic time series of increasing length. Since the actual efficiency of each algorithm depends to some extent on the character of the original time series, we considered uniform random noise (which has no structure and on average produces almost-balanced binary search trees), a Conway series (which has a quite rich structure and corresponds to a quite unbalanced tree), and a random walk series (which represents the more realistic scenario of a signal with both structure and noise). 

In the first case we observe the largest gap in computation time between the basic algorithm and the more efficient ones as it corresponds to the aforementioned average case where both algorithms (DC and the proposed one) significantly reduce the number of operations.
Such differences are more prominent in the computation of the horizontal visibility graph.

Additionally, in Figure~\ref{fig:speech_finance} we present a similar computational time analysis over real samples of speech (English language)~\cite{Garofolo93_TIMIT_NASA} and financial data~\cite{musmeci2017multiplex}. Figure~\ref{fig:speech_finance} is particularly interesting as it clearly shows a correlation between time computation and the time series structure (please note the different scale for time computation). Even though the time computation may differ, the DC and proposed method distribution seem to vary very little between data types in comparison to the relatively high spread observed for the basic algorithm. 

The horizontal visibility computation remains stable in both the DC and proposed method, and could potentially be considered independent of the data type given a time computation scaling factor. This behaviour was expected as the proposed method is fully defined by the aforementioned connectivity rules and has average-case time complexity $O(n \log n)$.

On the other hand, Figure~\ref{fig:speech_finance} suggests that the efficiency of the computation of natural visibility graphs is subject to wider fluctuations. The position of the maximum in the time series affects the efficiency of both the DC and the proposed method, as it will determine the number of additional visibility checks needed to obtain the natural visibility graph. 

An English speech time series will typically have its maximum somewhere towards the middle section of the signal (since we rarely tend to raise our voice at the end of our speech). Therefore the speech time series proposed codec will most probably produce an almost balanced binary search tree, yielding a time complexity of $O(n\log n)$. For this reason, one may observe a wider gap in computation time between the basic method and the faster alternatives for the speech data in Figure~\ref{fig:speech_finance} than for the financial time series.

\pagebreak
In terms of computation time, the proposed method and the DC one are closely related. They are both quicker than the basic implementation in both natural and horizontal visibility and they both present similar trends for increasing time series size (Figure~\ref{fig:series}). However, the proposed algorithm has proven to consistently be the quickest option for horizontal visibility graph computation. On the other hand, the DC algorithm in general does perform better than the proposed method for natural visibility computation. 
Even though at this point both DC and the proposed method seem equally good of an option for fast visibility computation, the presented algorithm has the additional property of allowing on-line assimilation of new data, which is something not easily achievable in either the basic approach or the DC algorithm.

The most straightforward way to asses the on-line functionality of the proposed method is to compare it with the equivalent off-line approach. In our case, it directly relates to the binary tree codec. Given a batch of new points to be added to the time series visibility analysis, in the off-line approach, the new batch is simply added to the time series itself and then the binary tree codec must be re-computed from scratch. In the proposed on-line approach, the next batch is encoded into its own binary tree that is then merged to the existing codec using the procedure detailed in Algorithm~\ref{Merge}. Note that the decoding step remains the same for the on-line and off-line approach, and so the comparison will essentially be between computing a codec from scratch (off-line) and merging two codecs into a single binary search tree (on-line). 

Figure~\ref{fig:merging} shows how much quicker the computation of the on-line method (codec for new data + merging) is in comparison to the computation time of the off-line approach (codec from scratch), for different time series and batch sizes. In particular, the on-line approach is always better if the new batch to be added is equal or bigger than the existing time series, especially for large time series.  

\newpage
\section{Conclusion}
The proposed visibility algorithm based on an encoder/decoder approach is, up to the authors' knowledge, the first efficient on-line algorithm to compute  visibility graphs. The analysis and the numerical experiments shown in the paper confirm that the proposed algorithm represents a substantial improvement over the state-of-the art for horizontal visibility computation, and is on par with the most efficient natural visibility algorithm (i.e. DC) available. Moreover, the procedure to assimilate new data by means of merging the corresponding binary search tree encoding into the existing tree allows for efficient on-line computation of visibility graphs, and represents a substantial speed-up with respect to the existing off-line algorithms.
This novel on-line capability broadens the applications for visibility graphs at no additional computational cost.

\bibliographystyle{abbrv}
\bibliography{bibtex/referencesMusic}

\end{document}